\documentclass{appolb}
\usepackage{graphicx}
\usepackage[utf8]{inputenc}
\usepackage[OT1]{fontenc}

\usepackage{amsfonts,amssymb}
\usepackage{lmodern} 

\newcommand{\const}{\mbox{\rm const}}

\newcommand{\osc}{\mbox{\rm osc}}
\newcommand{\var}{\mbox{\rm Var}}
\newcommand{\vol}{\mbox{\rm vol}}
\newcommand{\mod}{\mbox{\rm mod}}

\usepackage{bold-extra}

\newtheorem{definition}{Definition} 
\newtheorem{theorem}{Theorem} 
\newtheorem{proposition}{Proposition} 
\newenvironment{proof}{\noindent\small{\bf Proof:}}{{\hfill $\Box$}}

\begin{document}

\title{Fractality of certain quantum states}
\author{Daniel K. Wójcik\footnote{email: d.wojcik@nencki.edu.pl}
  \address{Nencki Institute of Experimental Biology of Polish Academy of Sciences, ul. Pasteura 3, 02-093 Warsaw, Poland}
  \address{Institute of Applied Psychology, Jagiellonian University, {\L}ojasiewicza 
     4, 30-348 Krak{\'o}w, Poland}
  \\[1em]
  Karol Życzkowski 
   \address{Institute of Theoretical Physics, Jagiellonian University, {\L}ojasiewicza 
     11, 30-348 Krak{\'o}w, Poland}
    \address{Center for Theoretical Physics, Polish Academy of Sciences, Al. Lotnik\'ow 32/46, 26-680 Warszawa, Poland}
}

\date{January 20, 2023}
\maketitle

\begin{abstract}
  Fractal structures appearing in solutions of certain quantum
  problems are investigated.  We prove the previously  
  announced results concerning the
  existence and properties of fractal states for the Schr\"odinger
  equation in the infinite one-dimensional well. 
  In particular, we show that for this problem there exist 
  solutions in the form of {\sl fractal quantum carpets}: the probability density $P(x,t)$ forms a fractal surface with dimension $D_{xy}$,
  while its cross-sections $P_t(x)$ and $P_x(t)$
   typically form fractal graphs with dimensions $D_x$
   and $D_t$ respectively, where  $D_{xy}=2+D_x/2$ and 
   $D_t=1+D_x/2$ (almost everywhere).
   
\end{abstract}
  
\section{Introduction}
\label{sec:intro}

Fractals are sets and measures of non-integer
dimension~\cite{Mandelbrot82da, Falconer90da}. They are good models of
phenomena and objects in various areas of science.  Their ubiquity in
dynamical systems theory as attractors, repellers, and attractor
boundaries is well-known~\cite{Ott93da,Falconer90da}. They are often
connected with non-equilibrium problems of growth~\cite{Meakin98da}
and transport~\cite{Gaspard98sa,Dorfman99s}. Fractal properties of
hydrodynamic modes were shown to be connected with transport
coefficients~\cite{Gilbert00s,Gaspard01s}. Fractal dimensions are used
in many nonlinear time series analysis
methods~\cite{wojcik01da,Kantz2004}

Fractals have also been found in quantum mechanics
\cite{La00,Ja14,La18,BI19}. For instance, quantum models related to
the problem of chaotic scattering often reveal fractal structures
\cite{Eckhardt88da,Ketzmerick96da,Casati00da}, relevant for quantum
transport \cite{KRHH13}.  Fractal structures play a prominent role in
studies of quantum dynamics of a reduced density operator
\cite{KKK22}.
A spectroscopic characterisation of the electronic wave function
inside a confined structure with fractal geometry was discussed in
\cite{KSF19}.  Quantum field theories in fractal spacetimes were also
analyzed \cite{Kroeger00da,Ak12} and fractal structures were reported
in models of quantum gravity \cite{AJW95,AJL05}.  Fractional calculus
was found useful to describe dynamics of quantum particles
\cite{Ta10}, while a Bohmian approach to quantum fractals was
presented in \cite{Sa05}.

It was also 
shown
that Schr\"odinger equation for the simplest non-chaotic potentials
admits fractal solutions~\cite{Berry96da}. 
The resulting probability distributions $P(x,t)$ as functions of space
and time, called {\sl quantum carpets} \cite{MSBB98,
  HRS99,BMS01,GKBR20}, reveal fractal
features~\cite{Berry96da,wojcik01db,GS03}. In this paper we
have two objectives. One is to present the
rigorous proofs of fractality of quantum states reported
in~\cite{wojcik01db}. The other is to illustrate a convenient method
of calculating dimensions of graphs of continuous functions introduced
by Claude Tricot~\cite{Tricot95da}.

\section{Methods}
\label{sec:methods}

\subsection{Box-counting dimension}
\label{sec:boxdim}

In this section we recall several equivalent definitions of
box-counting dimension, state a criterion for finding the dimension
of a continuous function of one variable and prove a connection
between the dimension of graph of function of $n$ variables and the
dimensions of its sections. All of this is known with the exception,
perhaps, of theorem~\ref{twie:multidim} which might be new. We
concentrate on the theory of box-counting dimension for graphs of
continuous functions of one variable. More general theory and a deeper
presentation can be found for instance
in~\cite{Falconer90da,Falconer97da,Tricot95da}. 

Let $A\subset {\mathbb R}^n$ be bounded. Consider a grid of
$n$-dimensional boxes of side $\delta$ 
\begin{equation}
  \label{eq:dboxes}
  [m_1\delta,(m_1+1)\delta] \times \dots \times [m_n\delta,(m_n+1)\delta].
\end{equation}
Let $N(\delta)$ be the number of these boxes covering the set $A$. It
is always finite because $A$ is bounded. 

\begin{definition}\label{defi:boxdim}
  {\em Box-counting dimension} of the set $A$ is the limit
  \begin{equation}
    \label{eq:boxdim}
    \dim_B (A) := \lim_{\delta\rightarrow 0}\frac{\ln N(\delta)}{\ln
    1/\delta}. 
  \end{equation}
  If the limit does not exist one considers {\em upper\/} and {\em lower
    box-counting dimensions\/} 
  \begin{eqnarray}
    \label{eq:boxupdown}
    \overline{\dim}_B (A) & := & \limsup_{\delta\rightarrow 0}
    \frac{\ln N(\delta)}{\ln 1/\delta},\label{eq:upboxdim}\\ 
    \underline{\dim}_B (A) & := & \liminf_{\delta\rightarrow0}
    \frac{\ln N(\delta)}{\ln 1/\delta}\label{eq:lowboxdim}
  \end{eqnarray}
  which always exist and satisfy
  \begin{equation}
    \overline{\dim}_B (A) \geq \underline{\dim}_B (A).
  \end{equation}
  The box-counting dimension exists if the upper and lower box-counting
  dimensions are equal.
\end{definition}

Several equivalent definitions 
are in use (see \cite{Mandelbrot82da,Falconer90da,Falconer97da,Tricot95da}
for a review). The most convenient definition to study the fractal
properties of graphs of continuous functions is given in terms
of $\delta$-variations~\cite{Tricot95da}. It is essentially a variant
of Bouligand definition~\cite{Bouligand28m}. We shall restrict our
attention to dimensions of curves being subsets of a plane.

Let $K_\delta(x)$ be a closed ball $\{y\in {\mathbb R}^2 :\, |x-y|\leq
\delta\}. $
\begin{definition}
  {\em Minkowski sausage\/} or {\em $\delta$-parallel body\/} of
  $A\subset{\mathbb R}^2 $ is
  \begin{eqnarray}
    A_\delta & := &\bigcup_{x\in A} K_\delta(x) \label{minsaus}\\
    & = & \{ y \in {\mathbb R}^2 :\, \exists x\in A,\; |x-y|\leq
    \delta\}. 
  \end{eqnarray}
\end{definition}
Thus the Minkowski sausage of $A$ is the set of all the points located
within~$\delta$ of~$A$. 

\begin{proposition}
  The {\em box-counting dimension\/} of a set $ A\subset {\mathbb
    R}^2$ satisfies 
  \begin{equation}
    \label{eq:pro1}
    \dim_B (A) = \lim_{\delta\rightarrow 0}\left(2-\frac{\ln
        V(A_\delta)}{\ln \delta}\right), 
  \end{equation}
  where 
  $ 
  V(\delta) = \vol^2 (A_\delta)
  $ 
  is the area of the Minkowski sausage of $A$.
\end{proposition}
\begin{proof}
  Every square from the $\delta$-grid containing $x\in A$ is included
  in $K_{\sqrt{2}\delta}(x)$. On the other hand, every closed ball of
  radius $\sqrt{2}\delta$ can be covered by at most 16 squares from
  the grid. Therefore
  \begin{equation}
    \label{eq:sausage}
    \delta^2 N(\varepsilon) \leq V(A_{\sqrt{2}\delta}) \leq 16 \delta^2
    N(\varepsilon) .
  \end{equation}
\end{proof}

Consider a continuous function on a closed interval $f:[a,b]
\rightarrow {\mathbb R}$. Its graph is a curve in the plane. To find
its box-counting dimension estimate the number of boxes $N(\delta)$
intersecting the graph.
Choose column $\{(x,y) : x\in [n\delta,(n+1)\delta]\}$. Since the
curve is continuous the number of the boxes in this column
intersecting the graph of $f$ is at least
\begin{equation}
  \label{eq:oscn}
  \left[\sup_{x\in[n\delta,(n+1)\delta]} f(x) -
    \inf_{x\in[n\delta,(n+1)\delta]} f(x)\right] /\delta
\end{equation} 
and no more than the same plus 2.  If $f$ was a record of a signal
then the difference between the maximum and minimum value of $f$ on
the given interval quantifies how the signal oscillates on this
interval. That's why it is called {\em $\delta$-oscillation}.
\begin{definition}
  {\em $\delta$-oscillation} of $f$ at $x$ is 
  \begin{eqnarray}
    \osc_\delta(x)(f) & := & \sup_{|y-x|\leq\delta} f(y) -
    \inf_{|y-x|\leq\delta} f(y) \\
    & = & \sup\{|f(y)-f(z)| \, : \,
    y,z \in [a,b]\cap [x-\delta,x+\delta]\}.
  \end{eqnarray}
\end{definition}
We will skip $(f)$ if it is clear from the context which function we
consider. 

From (\ref{eq:oscn}) we obtain the following estimate on the total
number of boxes covering the graph of $f$:
\begin{equation}
  \sum_{m=1}^M \osc_{\delta/2}(x_m)/\delta \leq N(\delta)
  \leq 2M+\sum_{m=1}^M \osc_{\delta/2}(x_m)/\delta ,
\end{equation}
where $x_m=a+(m-1/2)\delta$ is the middle of the $m$--th column from
the cover of the graph and $M=\lceil\frac{b-a}{\delta}\rceil$ is the
number of columns in the cover ($\lceil x \rceil$ stands for the
smallest integer greater or equal to $x$). Thus 
\begin{equation}
  N(\delta) \approx M \, \overline{\osc}_{\delta/2}/\delta.
\end{equation}
If the graph of $f$ has the box-counting dimension $D$, $N(\delta)$
scales as $\delta^{-D}$. This implies the following scaling of the
oscillations 
\begin{equation}
  \overline{\osc}_{\delta/2}\approx N(\delta)
  \delta/M
  \propto \delta^{2-D}.
\end{equation}
We have thus suggested a connection between the box-counting
dimension of the graph and the scaling exponent of the average
oscillation of the function $f$.

\begin{definition}
  {\em $\delta$-variation} of function $f$ is
  \begin{eqnarray}
    \label{eq:var}
    \var_\delta (f) & := &\int_a^b \osc_\delta (x)(f) \,dx \\
    & =: & (b-a)  \overline{\osc}_\delta (f).
  \end{eqnarray}
\end{definition}
Geometrically, variation is the area of the set scanned by the graph
of $f$ moved horizontally $\pm\delta$ and truncated at $x=a$ and
$x=b$, thus it is a kind of Minkowski sausage constructed
with horizontal intervals of length $2\delta$. This observation
leads to a convenient technique for calculating dimensions.

\begin{theorem}
  \label{twierdze:tricot}
  Let $f(x)$ be a non constant continuous function on $[a,b]$, then 
  \begin{equation}
    \label{eq:vardim}
    \dim_B \mbox{\rm graph} f = \lim_{\delta\rightarrow0}
    \left(2-\frac{\ln \var_\delta(f)}{\ln \delta}\right).
  \end{equation}
\end{theorem}
The proof consists of showing equivalence of $\var_\delta(f)$ with the
Minkowski sausage and follows from inequality (\cite{Tricot95da},
p. 130--132, 148--149) 
\begin{equation}
  \label{eq:varsaus}
  \var_\delta(f) \leq V(A_\delta) \leq c \var_\delta(f) ,
\end{equation}
where 
\begin{eqnarray*}
  A & = & \mbox{\rm graph} f\\
  c & = & c_1+c_2/s \\
  s & = & \left[\sup_{x\in[a,b]} f(x) - \inf_{x\in[a,b]} f(x)\right].
\end{eqnarray*}
This is where the assumption of non-constancy of $f$ comes in. 
Derivation of~(\ref{eq:varsaus}) is not difficult but rather lengthy
and will be omitted.

This theorem is the main tool to prove Theorem~\ref{twierdze:well}. In
order to find the dimensions we will look for estimates of
$\delta$-variation. They will usually take the following form:
\begin{proposition} \label{propo:estdim}
  \
  \begin{enumerate}
  \item $\osc_\delta(x) f(x) \leq c \delta^{2-s} \Rightarrow
    \dim_B \mbox{\rm graph} f \leq s$.
  \item $W:=\int_a^b |f(x+\delta)-f(x-\delta)| dx \geq c \delta^{2-s}
    \Rightarrow  \dim_B \mbox{\rm graph} f \geq s$.
  \end{enumerate}
\end{proposition}
\begin{proof}
  \
  \begin{enumerate}
  \item $\var_\delta f = \int_a^b \osc_\delta(x) (f) dx \leq (b-a) c
    \delta^{2-s}$ .
  \item $\osc_{2\delta}(x) f \geq |f(x+\delta)-f(x-\delta)| \Rightarrow
    \var_\delta f \geq (b-a) c (\delta/2)^{2-s}$.
  \end{enumerate}
\end{proof}

To prove the last point of theorem~\ref{twierdze:well} we need to know
what is the dimension of the graph of $f:{\mathbb R}^n\rightarrow
{\mathbb R}$ given all the dimensions of its one-variable
restrictions.

\begin{theorem} \label{twie:multidim}
  Let $  f \in {\cal C}^0([a_1,b_1]\times\dots\times[a_n,b_n])$.
  For every point $x  =  (x^1,\dots,x^n)\in
  [a_1,b_1]\times\dots\times[a_n,b_n]$ define 
  $ \tilde{x}^i := (x^1,\dots,x^{i-1}, x^{i+1},\dots,x^n)$.
  Then 
  \begin{equation}
    f_i[\tilde{x}^i_0](x^i) := f(x^1_0, \dots, x^{i-1}_0, x^i,x^{i+1}_0,
    \dots, x^n_0) 
  \end{equation}
  is a restriction of $f$ to a line parallel to $i$-th axis going
  through $x_0$ and $f_i[\tilde{x}^i_0] \in {\cal C}^0([a_i,b_i])$.
  \begin{enumerate}
  \item If $\forall x : \osc_\delta f_i[\tilde{x}^i] \leq c_i \delta^{H_i}$ then   
    $
    \dim_B \mbox{\rm graph} f(x^1,\dots,x^n)
    \leq n+1-\min\{H_1,\dots,H_n\}.
    $
  \item If $\var_\delta f_i[\tilde{x}^i_0] \geq c_i \delta^{H_i}$ for
    a dense set $ \tilde{x}^i_0\in A \subset \overline{A} = [a_1,b_1]
    \times \dots \times [a_{i-1},b_{i-1}] \times  [a_{i+1},b_{i+1}]
    \times \dots \times [a_n,b_n]$ 
    then 
    $
      \dim_B \mbox{\rm graph} f(x^1,\dots,x^n)
      \geq  n+1-\min\{H_1,\dots,H_n\}.
    $ 
  \item If all of the above conditions are satisfied then
    \begin{eqnarray*}
      \dim_B \mbox{\rm graph} f(x^1,\dots,x^n)
      & = & n+1-\min\{H_1,\dots,H_n\}\\
      & = & n-1+\max\{s_1,\dots,s_n\} 
    \end{eqnarray*}  
    where 
    \[
    s_i= \sup_{\tilde{x}^i}\dim_B \mbox{\rm graph}
    f_i[\tilde{x}^i](x^i).
    \]
  \end{enumerate}
\end{theorem}

In other words, the strongest oscillations along any direction
determine the box-counting dimension of the whole $n+1$-dimensional
graph.

\begin{proof}
  We will show the theorem for $n=2$ for notational simplicity.
  Generalization to arbitrary $n$ is immediate.  Let
  $f:[a_1,b_1]\times [a_2,b_2] \rightarrow {\mathbb R}$.  Divide the
  domain into squares $X_i\times Y_j$ of side $\delta$.  This gives
  rise to $K$ columns $A_{ij}$ of $\delta$-grid in ${\mathbb R}^3$,
  $1\leq\frac{K\delta^2}{(b_1-a_1)(b_2-a_2)} \leq 2$.  
  \begin{enumerate}
  \item The number of $\delta$-cubes
    having a common point with the graph of $f$ in column $A_{ij}$ is not
    greater than $(\sup_{A_{ij}} f - \inf_{A_{ij}} f)/\delta +2$. But 
    \begin{eqnarray*}
      |f(x_1,y_1)-f(x_2,y_2)| & = &|f(x_1,y_1) - f(x_1,y_2) +
      f(x_1,y_2) - f(x_2,y_2)|\\
      & \leq &|f(x_1,y_1) - f(x_1,y_2)| +
      |f(x_1,y_2) - f(x_2,y_2)|.
    \end{eqnarray*}
    Therefore
    \begin{eqnarray*}
      \sup_{A_{ij}} f - \inf_{A_{ij}} f & = & \sup_{(x_1,y_1),(x_2,y_2) \in
      A_{ij}} |f(x_1,y_1)-f(x_2,y_2)|\\
      & \leq &  \sup_{x\in X_i}\sup_{y\in Y_j} f(x,y) +
      \sup_{y\in Y_j}\sup_{x\in X_i} f(x,y) \\
      & \leq & \sup_{x\in X_i}\osc_{\delta/2}f_1[x] + \sup_{y\in
      Y_j}\osc_{\delta/2}f_2[y] \\ 
      & \leq & c \delta^{\min\{H_1,H_2\}}.
    \end{eqnarray*}
    Thus
    \begin{eqnarray*}
      \dim_B \mbox{\rm graph} f(x^1,x^2)
      & \leq & \lim_{\delta\rightarrow 0} \frac{\ln (K c
        \delta^{\min\{H_1,H_2\}}/\delta)}{\ln{1/\delta}} \\
      & \leq & 3-\min\{H_1,H_2\}.
    \end{eqnarray*}
  \item 
    Set $x\in X_i$. From~(\ref{eq:sausage}) and~(\ref{eq:varsaus}) it
    follows that the number $N_i(\delta)$ of $\delta$-cubes in 
    columns $A_{ij}$ covering the graph of $f_2[x](y)$ and the
    variation of $f_2$ satisfy 
    \begin{equation}
      \var_\delta f_2[x] \leq c \delta^2 N_i(\delta).
    \end{equation}
    Thus
    \begin{equation}
      N_i(\delta) \geq c \sup_{x\in X_i} \var_\delta f_2[x] /\delta^2
      \geq c \delta^{H_2-2}.
    \end{equation}
    Therefore the number $N(\delta)$ of boxes covering the whole graph
    of $f$ satisfies
    \begin{equation}
      N(\delta) \geq c \sum_{i=1}^M \sup_{x\in X_i} \var_\delta
      f_2[x] /\delta^2 \geq c_1 \delta^{H_2-3}.
    \end{equation}
    The same can be repeated for any direction thus
    \begin{equation}
      N(\delta) \geq  c_2 \delta^{\min\{H_1,H_2\}-3}.
    \end{equation}    
    
  \item An immediate corollary.
  \end{enumerate}
  Generalization to arbitrary $n$ is achieved by observing that
  $K\delta^n\approx \const$.
\end{proof}

  Another definition which has some convenient technical properties is
  the Hausdorff dimension \cite{Hausdorff19da,Edgar98da,Falconer85da},
  however it is often too difficult to calculate. For instance,
  as far as we know, there is still no proof that the Hausdorff
  dimension of the Weierstrass function is equal to its box-counting
  dimension. Thus in practice one usually uses the (upper)
  box-counting dimension. This is also our present approach. It is
  often assumed that the box-counting dimension and the Hausdorff
  dimension are equal. A general characterization of situations when
  this conjecture really holds is also lacking.

\subsection{Fractal functions}
\label{sec:fractalfunctions}

One of the oldest fractals is a graph of the Weierstrass function
\cite{Weierstrass72da,Edgar93da}:
\begin{equation}
  W(x)=\sum_{n=0}^\infty a^n \cos (b^n x\pi).
  \label{Weier}
\end{equation}
introduced as an example of everywhere continuous nowhere
differentiable function by Karl Weierstrass around 1872. Maximum range
of parameters for which the above series has required properties was 
found by Godfrey Harold Hardy in 1916~\cite{Hardy16da}, who also showed
that 
\begin{equation}
  \label{weierosc}
  \sup\{ |f({x})-f({y})| : |x-y|\leq \delta\} \sim \delta^H,
\end{equation}
where 
\[
H=\frac{\ln (1/a)}{\ln b}.
\]
From this it easily follows (see below) that the box-counting
dimension of the graph of the Weierstrass function $W(x)$ is 
\begin{equation}
  D_W= 2 + H = 2+\frac{\ln a}{\ln b} = 2-\left|\frac{\ln a}{\ln b}\right|.
\end{equation}
Functions whose graphs have non-integer box-counting dimension are
called {\em fractal functions}. Even though the box-counting dimension
of the Weierstrass function is easy to calculate~\cite{Tricot95da},
the proof that its Hausdorff dimension has the same value is still
lacking, as far as we know. Lower bounds on the Hausdorff dimension of
the graph were found by Mauldin~\cite{Mauldin86da,Mauldin86db}. Graphs
of random Weierstrass functions were shown to have the same Hausdorff
and box-counting dimensions for almost every distribution of
phases~\cite{Hunt98da}.

\section{Results}
\label{sec:results}

The construction of the Weierstrass function, eq.~(\ref{Weier}), can
easily be realized in quantum mechanics. Consider solutions of the
Schr{\"o}dinger equation
\begin{equation}
  i\partial_t \Psi(x,t) = -\nabla^2 \Psi(x,t)
  \label{Schrowell}
\end{equation}
for a particle in one-dimensional infinite potential well. The
general solutions satisfying the boundary conditions
$\Psi(0,t)=0=\Psi(\pi,t)$ have the form
\begin{equation}
  \Psi(x,t)=\sum_{n=1}^{\infty} a_n \sin (nx) e^{-i n^2 t},
  \label{well1}
\end{equation}
where
\begin{equation}
  a_n=\frac2\pi \int_0^{\pi} \!\! dx\, \sin (nx) \Psi(x,0).
  \label{wellcoeff}
\end{equation}

Weierstrass quantum fractals are wave functions of the form
\begin{equation}
  \Psi_M(x,t)=N_M \sum_{n=0}^M q^{n(s-2)} \sin(q^n x)
  e^{-iq^{2n} t},
  \label{weierwell}
\end{equation}
where $q=2,3,\dots$, $s\in (0,2)$.

In the physically interesting case of finite $M$ the wave function
$\Psi_M$ is a solution of the Schr\"odinger equation. The limiting
case
\begin{equation}
  \label{wellinf}
  \Psi(x,t):=\lim_{M\rightarrow\infty} \Psi_M(x,t)= N
  \sum_{n=0}^\infty q^{n(s-2)} \sin(q^n x) e^{-iq^{2n} t},
\end{equation}
with the normalization constant $N=\sqrt{\frac2\pi(1-q^{2(s-2)})}$, is
continuous but nowhere differentiable.  It is a weak solution of the
Schr\"odinger equation. 
Note that Eq.~(\ref{wellinf}) converges for
\( 
  |q^{s-2}|<1 \quad \equiv \quad s<2.
\) 
Since the probability density of wave function~(\ref{wellinf}) shows
fractal features for $s>0$ (see below), the interesting range of $s$
is $(0,2)$.

The main results announced in~\cite{wojcik01db} which we prove here are 
that not only the real part of the wave function $\Psi(x,t)$,
but also the physically important probability density $P(x,t) : =
|\Psi(x,t)|^2$ exhibit fractal nature. This is not obvious, because
$|\Psi(x,t)|^2$ is the sum of squares of real and imaginary part
having usually equal dimensions. One can easily show that the
dimension of the graph of a sum of functions whose graphs have the
same dimensions $D$ can be anything\footnote{Let
  $f_1$ and $f_2$ be functions with graphs having dimensions,
  respectively, $1\leq D_1<D_2\leq 2$. Let $g_1=f_1+f_2$,
  $g_2=f_1-f_2$. Then the box-counting dimension of both the graph of
  $g_1$ and $g_2$ is $D_2$, but the dimension of the graph of their
  sum $g_1+g_2=2f_1$ is $D_1\in[1,D_2[$.} from 1 to $D$.

Our main results are given by
\begin{theorem}
  \label{twierdze:well}
  Let $P(x,t)$ denote the probability density
  of a Weierstrass-like wave function 
  (\ref{wellinf}). Then
  \begin{enumerate}
  \item at the initial time $t=0$ the probability density,
    $P_0(x)=P(x,0)$, forms a fractal graph in the space variable
    (i.e.\ a\/ {\em space fractal\/}) of dimension $D_x=\max\{s,1\}$;
  \item the dimension $D_x$ of graph of $P_t(x)=P(x,t=const)$ does not
    change in time; 
  \item for almost every $x$ inside the well the probability density,
    $P_x(t)=P(x=\const,t)$, forms a fractal graph in the time variable
    (i.e.\ a\/ {\em time fractal\/}) of dimension $D_t(x)=D_t:=1+s/2$;
  \item for a discrete, dense set of points $x_d$, $P_{x_d}(t)=P(x_d,t)$
    is smooth and thus  $D_t(x_d)=1$;
  \item for even $q$ the average velocity $\frac{d\langle
      x\rangle}{dt}(t)$ is fractal with the dimension of its graph equal
    to $D_v=\max\{(1+s)/2,1\}$;
  \item the surface $P(x,t)$ has dimension $D_{xy}=2+s/2$.
  \end{enumerate}
\end{theorem}

The physical meaning of the above theorem has been discussed
in~\cite{wojcik01db}.  Here we only emphasize that to generate a
fractal wave function with exact mathematically rigorous fractal
features with infinite scaling properties an infinite energy is
required.  However, even a few terms in the series defining the
function (\ref{wellinf}) can lead to physically interesting effects.

Our proof of Theorem \ref{twierdze:well}
is based on the power-law behavior of the average $\delta$-oscillation
of the infinite double sum present in $P(x,t) = |\Psi(x,t)|^2$, see
Eq.~(\ref{eq:probdens}) in Appendix. Some fundamental concepts and
facts used in the proof 
are given in Section~\ref{sec:boxdim}. Calculations of probability
density and average velocity are provided in the Appendix. Positive
real constants are denoted by $c, c_1, c_2,\dots$

\section{Proof of Theorem~\ref{twierdze:well}}
\label{sec:proof}

\begin{proof}
\begin{enumerate}
\item[1.] {\em at the initial time $t=0$ the probability density,
    $P_0(x)=P(x,0)$, forms a fractal graph in the space variable
    (i.e.\ a\/ {\em space fractal\/}) of dimension $D_x=\max\{s,1\}$;}
\item[2.] {\em the dimension $D_x$ of graph of $P_t(x)=P(x,t=const)$ does not
    change in time. } \\
    
  We will show that
  for every fixed $t$ the graph of the probability
  density $|\Psi|^2$~(\ref{eq:probdens}) as a function of $x$ has 
  the box-counting dimension $s$. 
  \begin{enumerate}
  \item Fix $t$. Let
    \[
    P_n(x):=\sum_{k=0}^n q^{k(s-2)} \sum_{l=0}^k \sin q^l x \sin
    q^{k-l}x \cos(q^{2l} - q^{2(k-l)})t,
    \]
    $q=2,3,\dots$.  It is a smooth function whose derivative at every
    point satisfies 
    \begin{eqnarray*}
      |P_n'(x)| & \leq & 2 \sum_{k=0}^n q^{k(s-2)} \sum_{l=0}^k q^l
      |\cos q^l x \sin q^{k-l}x|  \\ 
      & \leq & 2\sum_{k=0}^n q^{k(s-2)} \frac{q^{k+1}}{q-1}
      \leq d_1(s,q) q^{n(s-1)},
    \end{eqnarray*}
    where            
    $$        
    d_1(s,q)=\frac{2q^s}{(q-1)(q^{s-1}-1)}.
    $$
    Let $\delta=q^{-n}$. Then 
    \begin{equation}                                
      \osc_\delta(x)P_n
      \leq  2 \delta  \sup_{x\in[0,\pi]} |P_n'(x)|
      \leq  2 d_1(s,q) \delta^{2-s}.
    \end{equation}
    On the other hand, for 
    \begin{eqnarray*}
      R_n(x)& := &P(x)-P_n(x) \\ 
      & = & \sum_{k=n+1}^\infty q^{k(s-2)} \sum_{l=0}^k \sin q^l x
      \sin q^{k-l}x \cos(q^{2l} - q^{2(k-l)})t 
    \end{eqnarray*}
    we have
    \[
    \osc_\delta(x)R_n \leq 2
    \sum_{k=n+1}^\infty q^{k(s-2)} (k+1)  
    \leq \frac{4q^{(n+1)(s-2)}n}{(1-q^{s-2})^2}.    
    \]
    Polynomial growth is slower than exponential, therefore
    for arbitrarily small $\varepsilon$ there is some $M$ such that
    \[ 
    \forall n>M : \quad n < (q^\varepsilon)^n.
    \] 
    This leads to the following estimate of the oscillation of
    $R_n$: 
    \[
    \osc_\delta(x)R_n \leq     
    d_2(s,q) \delta^{2-s-\varepsilon},
    \]
    where
    \[
    d_2(s,q)=\frac{4q^{s-2}}{(1-q^{s-2})^2}.
    \]
    Thus for all $x$ and $\delta=q^{-n}$, where $(\ln n)/n <
    \varepsilon\ln q$,  we have
    \begin{eqnarray*}
      \osc_\delta(x)P & \leq & \osc_\delta(x)P_n +
      \osc_\delta(x)R_n\\ 
      & \leq & (2d_1+d_2) \delta^{2-s-\varepsilon}.
    \end{eqnarray*}
    From Proposition~\ref{propo:estdim} it follows that 
    \[
    \dim_B {\rm graph}\,P_t(x) \leq 2- (2-s-\varepsilon)=s+\varepsilon
    \mathop{\longrightarrow}_{\varepsilon\rightarrow0} s.
    \]
  \item Fix $t$. Let $f(x)=P(x,t)$. We want to show that
    \[ 
    W:=\int_a^b |f(x+\delta)-f(x-\delta)| dx \geq c \delta^{2-s}.
    \] 
    Take $a=0$, $b=\pi$. 
    Notice that (we skip the normalization constant) 
    \begin{eqnarray}
      W & = & \int_0^\pi |f(x+\delta)-f(x-\delta)| dx
      \label{def.w}\\ 
      & = &  \int_0^\pi \left| \sum_{k=0}^\infty q^{k(s-2)}
        \sum_{l=0}^k \{\sin q^l(x+\delta) \sin q^{k-l}(x+\delta)
        +\right.\nonumber\\  
      & & \left. - \sin q^l(x-\delta) \sin q^{k-l}(x-\delta) \}
        \,a_{kl} \right| dx \nonumber\\ 
      & = & \int_0^\pi \left| \sum_{k=0}^\infty q^{k(s-2)}
        \sum_{l=0}^k \{ \cos q^lx \sin q^{k-l}x \sin q^l \delta \cos
        q^{k-l}\delta \} \,a_{kl}\right| \,dx , \nonumber
    \end{eqnarray}
    where
    \[ 
    a_{kl} = \cos (q^{2k}-q^{2l}) t.
    \] 
    Take $|h(x)|\leq 1$. Observe, that
    \begin{eqnarray}        
      \int_a^b \left|\sum_i f_i(x)\right| dx & \geq & \int_a^b
      \left|h(x)\right|\left|\sum_i f_i(x)\right| dx \nonumber\\ 
      \label{h(x)} & \geq & \left|\int_a^b \sum_i h(x)f_i(x) dx
      \right| \\ 
      & \geq & \left|\int_a^b h(x)f_k(x) dx \right| - \sum_{i\neq k}
      \left|\int_a^b h(x) f_i(x)dx \right|.  \nonumber 
    \end{eqnarray}
    One can interchange the order of summation and integration
    because $f(x)$ is absolutely convergent.  Let us take
    $\delta=q^{-N}$, $h(x) = \sin q^m x \cos q^n x$. After
    substitution in (\ref{def.w}) using (\ref{h(x)}) we obtain
    \begin{eqnarray*}
      W & \geq &\left|  \sum_{k=0}^\infty q^{k(s-2)} \sum_{l=0}^k
      \right. \sin q^{l-N} \sin q^{k-l-N} \cdot\\
      &&\cdot \left. \int_0^\pi \!\! dx\,\sin q^lx \cos q^{k-l}x
        \cos q^m x \cos q^n x \,a_{kl}
      \right| \\ 
      & = & \frac{\pi}{4} q^{(m+n)(s-2)} \left| \cos
        q^{m-N} \sin q^{n-N} \cos (q^{2(m+n)}-q^{2m}) t \right| \\ 
      & =: &  \frac{\pi}{4}\widetilde{W}
    \end{eqnarray*}         
    
    In what follows we will show three consecutive proofs that
    \[ 
    \exists c: \quad \widetilde{W} = q^{(m+n)(s-2)} \left| \cos
      q^{m-N} \sin q^{n-N} \cos (q^{2(m+n)}-q^{2m}) t \right| \geq c
    q^{N(s-2)},
    \] 
    for $t=\frac{k\pi}{q^l}$, $t/\pi\in {\mathbb Q}$, and a general
    proof for arbitrary real $t$. We will take advantage of the fact
    that $q$ is integer.
    
    Let $N=m+n$. Then
    \[
    \widetilde{W} = q^{N(s-2)} \left| \cos
      q^{m-N} \sin q^{-m} \cos (q^{2N}-q^{2m}) t \right|.
    \]
    It is enough to consider $t\in[0,\pi[$.
          
    \begin{enumerate}
    \item Let $t=\frac{k\pi}{q^l}$. Take $m$ such that
      \(
      2m\geq l
      \),
      for instance $m=l$. Then 
      \[
      (q^{2N}-q^{2m})\frac{k\pi}{q^l} =  (q^{2N-l}-q^l)k\pi,
      \]
      therefore 
      \[
      \left| \cos\left[ (q^{2N}-q^{2l})
          \frac{k\pi}{q^l}\right]\right| = 1. 
      \]
      We also have $\sin q^{-l}=\const$ and 
      \[
      \cos 1 = \cos q^0 \leq \cos q^{l-N} \leq \cos q^{-\infty} = \cos
      0.
      \]
      This means that for $t=\frac{k\pi}{q^l}$, for sufficiently
      large $N$ 
      \[
      \widetilde{W} \geq q^{N(s-2)} \cos 1 \cdot \sin q^{-l} \cdot
      \cos 0 = \const\  q^{N(s-2)}.
      \]
      
    \item
      Let $t/\pi\in {\mathbb Q}$.  For the next two proofs let us
      write $t/\pi$ in $q$ basis
      \begin{equation}
        \label{eq:expansion}
        \frac{t}{\pi} = \frac{a_1}{q} +\frac{a_2}{q^2} +\frac{a_3}{q^3}
        +\dots = \sum_{k=1}^\infty \frac{a_k}{q^k},
      \end{equation}                
      where $a_k\in\{0,1,\dots,q-1\}$.
      Therefore 
      \begin{eqnarray}
        \lefteqn{\cos[(q^{2N}-q^{2m}) t]} \nonumber\\
        & = & \cos [\pi (
        q^{2N-1}a_1 + q^{2N-2}a_2 + \dots + a_{2N} + q^{-1}a_{2N+1}
        + \dots + \nonumber\\  
        && +q^{2m-1}a_1 + q^{2m-2}a_2 + \dots + a_{2m} + q^{-1}a_{2m+1}
        + \dots)] \nonumber\\
        & = & \cos \left[\pi \left (\frac{a_{2N+1}-a_{2m+1}}{q} +
            \frac{a_{2N+2}-a_{2m+2}}{q^2} +                                
            \dots \right)\right]. \label{eq:cosine}
      \end{eqnarray}
      If we could only choose $m$ so that the first two terms in
      this series cancel out, we would have a lower estimate on the
      cosine, because in this case 
      \[
      \left|\frac{a_{2N+3}-a_{2m+3}}{q^3} + \dots \right| \leq
      (q-1)\left(\frac{1}{q^3}+ \frac{1}{q^4}+ \dots \right) =
      \frac{1}{q^2}. 
      \]
      Thus 
      \[
      \cos[(q^{2N}-q^{2m}) t] \geq \cos(\pi/q^2) \geq \cos \pi/4.
      \]
      Take a rational $t$. In any basis $q$ the
      expansion~(\ref{eq:expansion}) of $t/\pi$ is finite or
      periodic. Finite expansions have been treated in the previous
      point, thus we assume here the expansion of $t/\pi$ is
      periodic with period $T$. Thus for sufficiently large $n$ we
      have
      \[
      a_{n+T}=a_n
      \]
      so that $t/\pi$ can be written as
      \[
      t/\pi = 0.a_1 a_2 \dots a_K (a_{K+1} \dots a_{K+T}),
      \]
      where $(a_{K+1} \dots a_{K+T})$ means ``repeat $a_{K+1} \dots
      a_{K+T}$ periodically ad infinitum''. We shall now estimate
      $\widetilde{W}$ for this $t$. Every $N > K$ can be written as
      $N=K+pT+r$, where $p$ is natural or 0 and
      $r\in\{1,2,\dots,T\}$. If we now take $m=K+r$ then not only
      the first two but all the terms in~(\ref{eq:cosine}) cancel
      out, therefore
      \[
      \cos[(q^{2N}-q^{2m}) t] =1.
      \]
      Obviously, 
      \begin{eqnarray*}
        \sin q^{-m} & \geq & \sin q^{-(K+T)},\\
        \cos q^{m-N} & = & \cos q^{-pT} \geq \cos 1,
      \end{eqnarray*}
      Therefore 
      \[
      \widetilde{W} \geq q^{N(s-2)}\, \sin q^{-(K+T)} \cos 1  = \const\
      q^{N(s-2)}. 
      \]
      
    \item
      General case of an arbitrary $t/\pi\in [0,1]$. Its expansion in
      $q$ basis is given by~(\ref{eq:expansion}).  Let $A$ be the set
      of all the two element sequences with elements from the set
      $\{0,1,\dots,q-1\}$. Thus
      \[
      A=\{\{0,0\}, \{0,1\}, \dots, \{0,q-1\}, \{1,0\}, \dots,
      \{q-1,q-1\} \} ,
      \]
      we write $A_{k,l}:=\{k,l\},\, k,l \in \{0,1,\dots,q-1\}$.
      Consider all the pairs of consecutive $q$-digits of $t/\pi$ of
      the form  
      \begin{equation}
        \{a_{2m+1}, a_{2m+2} \}, \label{eq:pairs}
      \end{equation}
      i.e.\ $\{a_1, a_2 \}, \{a_3, a_4 \},$ etc. Every such pair is
      equal to some $A_{k,l}$. Let $N_{k,l}$ be the first such $m$
      for which 
      \[
      A_{k,l}=\{a_{2m+1}, a_{2m+2} \}.
      \]
      If $A_{k,l}$ for given $k,l$ doesn't appear in the sequence of
      all the pairs~(\ref{eq:pairs}) we set $N_{k,l}=0$. 
      Let
      \[
      M = \sup_{k,l} N_{k,l}. 
      \]
      Thus if $n>M$ the sequence $\{a_{2n+1}, a_{2n+2}\}$ has appeared
      at least once among the pairs $\{a_1, a_2 \}, \{a_3, a_4
      \},\dots, \{a_{2M+1}, a_{2M+2} \}$. Therefore, for every $N>M$
      we can find such an $m\in{1,2,\dots,M}$ that
      \[
      \left|\cos\left[(q^{2N}-q^{2m})\frac{t}{\pi} \pi\right] \right|
      \geq \cos\frac{\pi}{q^2} \geq \cos \frac{\pi}{4} =
      \frac{\sqrt{2}}{2}. 
      \]
      Also
      \begin{eqnarray*}
        \sin q^{-m} & \geq & \sin q^{-M},\\
        \cos q^{m-N} & \geq & \cos q^{M-N} \geq \cos 1,
      \end{eqnarray*}
      which leads to
      \[
      \widetilde{W} \geq q^{N(s-2)} \frac{\sqrt{2}}{2}\sin q^{-M}
      \cos 1 = \const\ q^{N(s-2)}. 
      \]
    \end{enumerate}
    
    We have thus shown that for every $t$, for natural $q$ and for
    $\delta=q^{-N}$  
    \[
    W \geq \const\ \delta^{2-s},
    \]
    therefore (proposition~\ref{propo:estdim})
    \[
    \dim_B {\rm graph}\,P_t(x) \geq 2- (2-s)=s.
    \]      
  \end{enumerate}
  
\item[3.] {\em For almost every $x$, $D_t(x)=\dim_B \mbox{\rm graph}
    P_x(t)=D_t:=1+s/2$}. \\
                                
  We will use the form~(\ref{eq:probdens2}) of the probability
  density. It is enough to analyze the dimension of 
  \[ 
  \widetilde{P}(t) := \sum_{c=1}^\infty q^{2c(s-2)} \sin q^c x
  \sum_{d=1}^c q^{-d(s-2)}\sin q^{c-d}x \cos
  [(q^{2c}-q^{2(c-d)})t].   
  \] 
  
  \begin{enumerate}     
  \item Let 
    \[
    P_n(t):= \sum_{c=1}^n q^{2c(s-2)} \sin q^c x \sum_{d=1}^c
    q^{-d(s-2)}\sin q^{c-d}x \cos[ (q^{2c}-q^{2(c-d)})t].
    \]
    Then 
    \begin{eqnarray*}
      |P_n'(t)| & = & \left| \sum_{c=1}^n q^{2c(s-2)} \sin q^c x
        \sum_{d=1}^c q^{-d(s-2)} \sin q^{c-d}x \cdot\right.\\
      & & \left. \cdot (q^{2c}-q^{2(c-d)}) \sin
        [(q^{2c}-q^{2(c-d)})t] \right| \\ 
      & \leq & \sum_{c=1}^n q^{2c(s-2)} \sum_{d=1}^c q^{-d(s-2)}
      (q^{2c}-q^{2(c-d)})\\ 
      & = & \sum_{c=1}^n q^{2c(s-2+1)} \sum_{d=1}^c q^{-d(s-2)}
      (1-q^{-2d})\\
      & = & \frac{q^{2-s}}{q^{2-s}-1} \left[ q^s \frac{q^{ns}-1}{q^s
          -1} -  q^{2(s-1)}\frac{q^{2n(s-1)}-1}{q^{2(s-1)}-1} \right]
      + \\ 
      & & - \frac{q^{-s}}{q^{-s}-1} \left[ q^{s-2}
        \frac{q^{n(s-2)}-1}{q^{s-2}-1} - q^{2(s-1)}
        \frac{q^{2n(s-1)}-1}{q^{2(s-1)}-1} 
      \right].
    \end{eqnarray*}
    
    Therefore, for $n$ large enough 
    \[
    |P_n'(t)| \leq c_1 q^{n \max\{s,2(s-1), s-2\}} = c q^{ns}.
    \]
    Let $\delta=q^{-\alpha n}$. Then $q^n=\delta^{-1/\alpha}$ and
    \[
    \osc_\delta(t) P_n \leq c_1 2 \delta q^{ns} = 2c_1
    \delta^{1-s/\alpha}.  
    \]
    
    Let 
    \[
    R_n(t):=\widetilde{P}(t)-P_n(t).
    \]
    Then 
    \begin{eqnarray*}
      \osc_\delta(t) R_n & \leq & 2 \sum_{c=n+1}^\infty q^{2c(s-2)}
      \sum_{d=1}^c q^{-d(s-2)}\\
      & \leq & \frac{2q^{2-s}}{q^{2-s}-1} \sum_{c=n+1}^\infty
      q^{2c(s-2)+c(2-s)}\\
      & = & c_2 \delta^{(s-2)/\alpha}.
    \end{eqnarray*}
    To obtain consistent estimate we must set 
    \[
    1-\frac{s}{\alpha} =\frac{2}{\alpha} -\frac{s}{\alpha},
    \]
    which gives $\alpha=2$. Thus
    \[
    \osc_\delta(t)\widetilde{P} \leq (2c_1+c_2) \delta^{1-s/2}.
    \]
    
  \item Now we want to show that
    \[    
    W = \int_a^b \!\! dt\,
    |\widetilde{P}(t+\delta) - 
    \widetilde{P}(t-\delta)|  \geq c \delta^{1-s/2}.
    \]
    Set $a=0, b=2\pi$ for convenience. Then 
    \begin{eqnarray*}
      W & = & \int_0^{2\pi} \!\! dt \,\left| \sum_{c=1}^\infty q^{2c(s-2)}
        \sin q^c x \sum_{d=1}^c q^{-d(s-2)}\sin q^{c-d}x \right.\cdot\\
      & & \left.\cdot\left\{\cos [(q^{2c}-q^{2(c-d)})(t+\delta)]- \cos
          [(q^{2c}-q^{2(c-d)})(t-\delta)]\right\}\right| \\
      & = & \int_0^{2\pi} \!\! dt \, \left|\sum_{c=1}^\infty q^{2c(s-2)}
        \sin q^c x \sum_{d=1}^c q^{-d(s-2)}\sin q^{c-d}x \right. \cdot\\
      & & \cdot\left.\left\{ -2 \sin [(q^{2c}-q^{2(c-d)})t]
          \sin [(q^{2c}-q^{2(c-d)})\delta]\right\}\right| .
    \end{eqnarray*}
    Using our standard arguments we multiply the integrand by a
    suitable function smaller or equal to 1:
    \begin{eqnarray*}
      W & \geq &  \int_0^{2\pi} \!\!dt \,|h(t)||\widetilde{P}(t+\delta) -
      \widetilde{P}(t-\delta)| \\
      & \geq & \left| \int_0^{2\pi} \!\!dt \,h(t)[\widetilde{P}(t+\delta) -
        \widetilde{P}(t-\delta)]\right|.
    \end{eqnarray*}
    We choose $h(t) = \sin[(q^{2c}-q^{2(c-d)})t]$ and set
    $\delta=q^{-2N}$. It follows that 
    \[
    W \geq 2\pi q^{(2c-d)(s-2)} \left|\sin (q^c x) \sin (q^{c-d}x) \sin
      [(q^{2c}-q^{2(c-d)})q^{-2N}]\right|. 
    \]
    We now want to show that for almost all $x$ 
    \[
    W \geq c_3 \delta^{1-s/2} = c_3 q^{N(s-2).}
    \]    
    Set $2c-d=N$. Then 
    \begin{equation}
      \label{ineq:w}
      W \geq 2\pi q^{N(s-2)} \left|\sin (q^c x) \sin (q^{N-c}x) 
        \sin [q^{2(c-N)}-q^{-2c}]\right|. 
    \end{equation}
    Thus it is enough to bound 
    \begin{equation}
      \left| \sin (q^c x) \sin (q^{N-c}x) \sin [q^{2(c-N)}-q^{-2c}]
      \right| \label{eq:tobound}
    \end{equation}
    from below. 
    
    Choose rational $x/\pi$. All the rational numbers in a given
    basis $q$ have one of the two possible forms: finite or
    periodic. In the first case ($x/\pi=k/q^l$) we cannot find the
    lower bound on~(\ref{eq:tobound}). We cannot succeed, because
    at these points the function $P_x(t)$ is smooth (cf.\ the
    next point of the proof). 
    
    The other case means that $x/\pi$ can be written as
    \[
    x/\pi = 0.a_1 a_2\dots a_K (a_{K+1}\dots a_{K+T}),
    \]
    where again $(a_{K+1}\dots a_{K+T})$ denotes the periodic
    part. Therefore, for every $N>K$, $q^n x \,\mod\, \pi$ can take only
    one of $T$ values: $q^{K+1} x \,\mod\, \pi, \dots, q^{K+T} x
    \,\mod\,\pi$. Let us take $c=1$, $N>K$. Then $|\sin (q^c x)|= |\sin
    (qx)|>0$ and is a 
    constant. $|\sin (q^{N-1} x)|$ takes one of $T$ values, none of
    which is $0$, therefore it is always bounded from below by
    \[
    \inf_{l=1,2,\dots,T} |\sin (q^{K+l} x)| > 0.
    \]
    Also the last term can be bounded:
    \[
    |\sin(q^{-2(N-c)}-q^{-2c})| \geq \sin (q^{-2} - q^{-2(N-1)})
    \geq \sin q^{-3}
    \]
    for $N\geq3$.
    Thus for rational $x/\pi$ with periodic expansion in $q$
    \[
    W \geq  c_3 q^{N(s-2)} ,
    \]
    where $c_3=2\pi |\sin (qx)\sin (q^{-3} x)|\inf_{l=1,2,\dots,T}
    |\sin (q^{K+l} x)| $.
    
    Consider now irrational $x/\pi$. Inequality (\ref{ineq:w}) for
    $c=N$ takes the form
    \begin{eqnarray}
      W & \geq & 2\pi q^{N(s-2)} \left|\sin (q^N x) \sin x 
        \sin [1-q^{-2N}]\right| \nonumber\\
      & \geq & c q^{N(s-2)} \left|\sin (q^N x)\right|,
      \label{ineq:w1} 
    \end{eqnarray}
    for $N\geq 2$.
    Instead of showing it can be bounded from below we will use it
    to prove that for almost every $x$ 
    \begin{equation}
      \dim_B \mbox{\rm graph} P_x(t) \geq 1+s/2.
    \end{equation}
    
    Let 
    \begin{equation}
      \label{eq:Renyi}
      x_n := q^n (x/\pi) \, \mod \, 1.
    \end{equation}
    Let 
    \[
    F_N^\alpha := \left\{ x \, : \, \exists \,n\geq N  \quad
      \left(x_n\leq \frac{1}{q^{N\alpha}}\right) \vee \left(1-x_n\leq
        \frac{1}{q^{N\alpha}}\right)\right\},
    \]
    $\alpha\in [0,1]$.
    Let 
    \[
    F_\infty^\alpha := \bigcap_{N=1}^\infty F_N^\alpha 
    \]
    Clearly, 
    \[
    F_N^\alpha \supset F_{N+1}^\alpha \supset F_{N+2}^\alpha \dots
    \]
    Since Renyi map (\ref{eq:Renyi}) preserves the Lebesgue measure
    we have 
    \begin{equation}
      \mu(F_N^\alpha) \leq 2\left(\frac{1}{q^N\alpha} +
        \frac{1}{q^{N+1}\alpha} + \frac{1}{q^{N+2}\alpha} + \dots\right) =
      \frac{2q}{q-1} \frac{1}{q^{N\alpha}}.
    \end{equation}
    Therefore
    \begin{equation}
      0 \leq \mu(F_\infty^\alpha) \leq \inf_N \mu(F_N^\alpha) = 0
    \end{equation}
    
    It follows that for almost every $\{x_n\}$ 
    \begin{equation}
      \lim_{n\rightarrow\infty} \frac{\ln |\sin x_{n}|}{n} \geq
      \lim_{n\rightarrow\infty} \frac{\ln q^{-n\alpha}}{n} \geq 
      \lim_{n\rightarrow\infty} \frac{q^{-n\alpha}}{2n} \geq
      -\alpha \ln q.
    \end{equation}    
    Thus for every $\alpha>0$ and for almost every
    $x/\pi\in[0,1]$ we have
    \begin{eqnarray}
      \dim_B \mbox{\rm graph} P_x(t) & = & \lim\left(2 - \frac{\ln
          \var_\delta P_x(t)}{\ln q^{-2N}}\right) \\
      & \geq & 2+\lim\left(\frac{\ln \var_\delta P_x(t)}{2N\ln q}\right) 
    \end{eqnarray}
    But 
    \begin{equation}
      \var_\delta P_x(t) \geq W 
    \end{equation}
    therefore from (\ref{ineq:w1})
    \begin{eqnarray}
      \dim_B \mbox{\rm graph} P_x(t) & \geq & 2+\lim\left(\frac{\ln c +
          N(s-2)\ln q + \ln |\sin (x_n \pi)|}{2N\ln q}\right) \\
      & = & 1+s/2 + \lim\left(\frac{\ln |\sin (x_n \pi)|}{2N\ln q}\right) \\ 
      & \geq & 1+s/2 - \alpha.
    \end{eqnarray}
    But $\alpha$ is arbitrary, thus
    \begin{equation}
      \dim_B \mbox{\rm graph} P_x(t) \geq 1+s/2.
    \end{equation}        
  \end{enumerate}
                   
\item[4.] {\em For a discrete, dense set of points $x_d$, $D_t(x_d)
    = \dim_B \mbox{\rm graph} P_{x_d}(t)=1$}.\\
  
  Let $x_{k,m}=\frac{m\pi}{q^k}$, where $k\in{\mathbb N}$,
  $m=0,1,\dots,q^k-1$. The set $\{x_{k,m}\}$ is dense in $[0,1]$. At
  these points $\Psi(x_{k,m},t)$ is a sum of a finite number of
  terms: 
  \[
  \Psi\left(\frac{m\pi}{q^k},t\right)=
  \sqrt{\frac2\pi\left(1-\frac{1}{q^{2(2-s)}}\right)}
  \sum_{n=0}^{k-1} q^{(s-2)n} \sin (q^{n-k} m\pi )\,e^{-iq^{2n}t}.
  \]
  Therefore,
  \[
  \dim_B \mbox{\rm graph}
  \left|\Psi\left(\frac{m\pi}{q^k},t\right)\right|^2 = 1.
  \]
  
\item[5.] {\em For even $q$ the average velocity $\frac{d\langle
      x\rangle}{dt}(t)$ is fractal with the dimension of its graph
    equal to $D_v=\max\{(1+s)/2,1\}$}.\\
  
  Heuristically, this is rather obvious because
  \[  
  \frac{d\langle x\rangle}{dt}(t) \approx 
  \sum_{k=1}^\infty \frac{q^{k(s-1)}}{q^{2k}} \sin q^{2k} t 
  = \sum_{k=1}^\infty q^{2k(s-3)/2} \sin q^{2k} t.
  \]
  Thus the average velocity is essentially a Weierstrass-like
  function and the dimension of its graph should be 
  \[
  2-\frac{3-s}{2} =\frac{1+s}{2}. 
  \]

  It is enough to consider 
  \[
  W(t) := \sum_{k=1}^\infty \frac{q^{k(s-1)}}{q^{2k}-1} \sin
  (q^{2k}-1)t. 
  \]
  \begin{enumerate}
  \item Let                                
    \[
    W_n(t) := \sum_{k=1}^n \frac{q^{k(s-1)}}{q^{2k}-1} \sin (q^{2k}-1)t.
    \]                              
    Set $\delta=q^{-\alpha n}$. Then                                
    \[ 
    |W_n'(t)| = \left|\sum_{k=1}^n q^{k(s-1)} \cos
      (q^{2k}-1)t\right| 
    \leq \sum_{k=1}^n q^{k(s-1)} 
    \leq c_1 \delta^{(1-s)/\alpha},
    \]    
    where
    \[
    c_1 = \frac{q^{s-1}}{q^{s-1}-1}.
    \]
    Therefore
    \[      
    \osc_\delta(t)W_n \leq 2 c_1
    \delta^{(1-s)/\alpha}\delta = 2 
    c_1 \delta^{1+(1-s)/\alpha}.
    \] 
    Now, for
    \[ 
    P_n(t):=W(t)-W_n(t)
    \] 
    we have
    \begin{eqnarray*}
      |P_n(t)| & = & \left|\sum_{k=n+1}^\infty \frac{q^{k(s-1)}}{q^{2k}-1} \sin
        (q^{2k}-1)t\right|
      \leq  \sum_{k=n+1}^\infty \frac{q^{k(s-1)}}{q^{2k}-1}\\ 
      & \leq & 2 \sum_{k=n+1}^\infty \frac{q^{k(s-1)}}{q^{2k}}
      = c_2 \delta^{-\frac{s-3}{\alpha}},
    \end{eqnarray*}
    where
    \[ 
    c_2=\frac{2q^{s-3}}{1-q^{s-3}}.
    \] 
    Thus
    \[ 
    \osc_\delta(t)P_n\leq 2 c_2 \delta^{\frac{3-s}{\alpha}}.
    \] 
    To obtain consistent estimate for both $P_n$ and $W_n$ we must
    set 
    \[
    1+\frac{1-s}{\alpha}  = \frac{3-s}{\alpha},
    \]
    thus $\alpha=2$ and $\delta=q^{-2N}$. Therefore
    \[
    \osc_\delta(t)W \leq \osc_\delta(t)W_n+\osc_\delta(t)P_n
    \leq 2 (c_1+c_2)
    \delta^{2-\frac{s+1}{2}}.
    \]
    
  \item Consider
    \begin{eqnarray*}
      \lefteqn{ \int_a^b |W(t+\delta)-W(t-\delta)| dt}\\ 
      & = & \int_a^b \!\!dt \, \left|\sum_{k=1}^\infty
        \frac{q^{k(s-1)}}{q^{2k}-1} \cos(q^{2k}-1)t \,\sin 
        (q^{2k}-1)\delta\right| \\ 
      & \geq & \left| \int_a^b\!\!dt \, h(t) f_N(t) \right| -
      \sum_{k\neq N} \left| \int_a^b \!\!dt \,h(t) f_k(t)\right| ,  
    \end{eqnarray*}
    where
    \[ 
    f_k(t) = \frac{q^{k(s-1)}}{q^{2k}-1}\cos(q^{2k}-1)t \, \sin
    (q^{2k}-1)\delta.  
    \] 
    Let $h(t)=\cos (q^{2N}-1)t$, $\delta=q^{-2N}$. Then
    \begin{eqnarray*}
      \lefteqn{\left| \int_a^b h(t) f_N(t) dt\right|}\\ 
      & = & \frac{q^{N(s-1)}}{q^{2N}-1} \sin(1-q^{-2N}) \int_a^b
      \!\!\cos^2(q^{2N}-1)t\,dt\\  
      & \geq &\frac{q^{N(s-1)}}{q^{2N}} \sin \frac{\pi}{6} \int_a^b
      \!\! \cos^2(q^{2N}-1)t\,dt\\ 
      & \geq & \frac12 q^{N(s-3)} \left[\frac{b-a}{2} +
        \frac{\sin(2b(q^{2N}-1))-\sin(2a(q^{2N}-1))}{4(q^{2N}-1)}\right]\\ 
      & \geq & \frac12\delta^{\frac{3-s}{2}} \left[\frac{b-a}{2} -
        \frac{2\cdot2}{4\cdot q^{2N}}\right]\\  
      &= &\frac12\delta^{\frac{3-s}{2}} \left[\frac{b-a}{2} -
        \delta\right]\\  
      & \geq &\frac18\delta^{\frac{3-s}{2}}(b-a).
    \end{eqnarray*}
    On the other hand
    \begin{eqnarray*}
      \lefteqn{\left| \int_a^b h(t) f_k(t) dt\right|}\\ 
      & = &\frac{q^{k(s-1)}}{q^{2k}-1} \sin\left[(q^{2k}-1)q^{-2N}
      \right] \int_a^b \!\! \cos(q^{2k}-1)t \cos(q^{2N}-1)t\,dt\\ 
      & \leq & \frac{q^{k(s-1)}}{q^{2k}-1} \left|
        \frac{\sin(b(q^{2N}-q^{2k})) -
          \sin(a(q^{2N}-q^{2k}))}{2(q^{2N}-q^{2k})} + \right.\\ 
      &&\left.\frac{\sin(b(q^{2N}+q^{2k}))-\sin(a(q^{2N}+q^{2k}))}{2(q^{2N}+q^{2k})}
      \right| \\ 
      & \leq
      & 2 q^{k(s-3)}\left[\frac{1}{|q^{2N}-q^{2k}|} +
        \frac{1}{q^{2N}+q^{2k}}\right]\\ 
      & \leq & 2 q^{k(s-3)}\left[\frac{1}{q^{2N}-q^{2(N-1)}} +
        \frac{1}{q^{2N}} \right] \\ 
      & \leq & 5
      \cdot q^{k(s-3)} \delta.
    \end{eqnarray*}
    Therefore
    \begin{eqnarray*}
      W & \geq & \frac18\delta^{(3-s)/2}(b-a) - \sum_k 5
      q^{k(s-3)} \delta\\ 
      & \geq & \frac18\delta^{(3-s)/2}(b-a) - 5
      \frac{q^{s-3}}{q^{s-3}-1}\delta. 
    \end{eqnarray*}
    But $\frac{3-s}{2}<1$, thus for large enough $N$ (small enough
    $\delta$) the first term dominates the other, therefore
    \[ 
    W \geq c \delta^{2-(1+s)/2},
    \] 
    with $c=\frac{b-a}{16}$, for example.
  \end{enumerate}
  From theorem~\ref{twierdze:tricot} it follows that
  \[ 
  D_v=\frac{1+s}{2}.
  \] 
  
\item[6.] {\em The surface $P(x,t)$ has dimension $D_{xy}=2+s/2$}.
  \\
  Setting $x$ or $t$ constant we have shown that oscillations are
  bounded by $c\delta^H$ where exponent $H$ is one of $1, s, s/2$.  We
  also showed the lower bound of variation is always $c\delta^H$ again
  with $H$ being one of $1, s, s/2$. What's more, there is a dense set
  of points $x$ for which $\var_\delta P_x(t) \geq c\delta^{s/2}$. One
  can take for instance all rational $x/\pi$ with periodic
  $q$-expansion. Thus from theorem~\ref{twie:multidim} we have
  \begin{equation}
    D_{xt}=1+\max\{D_x,D_t\}=2+\frac{s}{2}.
  \end{equation}            
\end{enumerate}
\end{proof}


\section{Acknowledgements}

We thank Iwo Bia{\l}ynicki-Birula for many enlightening comments
during our fruitful collaboration which culminated in the work ~\cite{wojcik01db},
where the results proven here where stated.


\appendix

\section{Auxiliary calculations}
\label{app:aux}

\subsection{\ Probability density}

Take the fractal wave function~(\ref{wellinf}),
\[
\Psi(x,t) = \sqrt{\frac2\pi\left(1-q^{2(s-2)}\right)}
\sum_{n=0}^\infty q^{n(s-2)} \sin(q^n x) e^{-iq^{2n} 
  t}.
\]
Let us calculate two useful forms of the probability density $P(x,t)$
\begin{eqnarray*}
  P(x,t) & = & |\Psi(x,t)|^2\\
  & = & \frac2\pi\left(1-q^{2(s-2)}\right) \sum_{m,n=0}^\infty
  q^{(m+n)(s-2)} \sin q^n x \sin q^m x\, e^{-i(q^{2n} - q^{2m})t}.
\end{eqnarray*}
Taking $k=m+n,l=n$ we obtain
\begin{eqnarray}
  P(x,t)& = & \frac2\pi\left(1-q^{2(s-2)}\right) \sum_{k=0}^\infty q^{k(s-2)}
  \sum_{l=0}^k \sin q^l x \sin q^{k-l}x \,e^{-i(q^{2l} - q^{2(k-l)})t},\nonumber\\
  & = & \frac2\pi\left(1-q^{2(s-2)}\right) \sum_{k=0}^\infty q^{k(s-2)}
  \sum_{l=0}^k \sin q^l x \sin q^{k-l}x \cos(q^{2l} -
  q^{2(k-l)})t,\qquad\quad. \label{eq:probdens} 
\end{eqnarray}
Substitute $c=m,\,d=m-n$ to arrive at
\begin{eqnarray}
  P(x,t) & = & \frac2\pi\left(1-q^{2(s-2)}\right) \sum_{m=0}^\infty
  \left\{ q^{2m(s-2)} \sin^2 q^m x + \right.\nonumber\\ 
  &&  \left.+ 2\sum_{n<m} q^{(m+n)(s-2)} \sin q^n x \sin q^m x \cos(q^{2m} -
    q^{2n})t, \right\}\nonumber\\
  &=&\frac2\pi\left(1-q^{2(s-2)}\right) 
  \left\{
    \sum_{m=0}^\infty q^{2m(s-2)} \sin^2 q^m x +\right.\nonumber\\
  &&\left.+2\sum_{c=1}^\infty\sum_{d=1}^c 
    q^{(2c-d)(s-2)} \sin q^c x \sin q^{c-d} x \cos (q^{2c}-q^{2(c-d)})t
  \right\}\nonumber\\
  &=& \frac2\pi \left(1-q^{2(s-2)}\right) 
  \left\{
    \sum_{m=0}^\infty q^{2m(s-2)} \sin^2 q^m x+\right.\nonumber\\
  && +2\sum_{c=1}^\infty q^{2c(s-2)} \sin q^c x \cdot\nonumber\\
  && \left.\cdot \sum_{d=1}^c q^{-d(s-2)}  \sin q^{c-d} x \cos \left[(q^2-1)
      \cdot   q^{2(c-d)}\sum_{a=0}^{d-1}q^{2a} \right]t \right\}\nonumber\\
  &=&\frac2\pi\left(1-q^{2(s-2)}\right) 
  \sum_{m=0}^\infty q^{2m(s-2)} \sin^2 q^m x+\nonumber\\
  &&+\frac4\pi\left(1-q^{2(s-2)}\right) \sum_{c=1}^\infty q^{2c(s-2)}
  \sin q^c x \cdot\nonumber\\
  &&\cdot \sum_{d=1}^c  q^{-d(s-2)}\sin q^{c-d}x \cos (q^2-1)
  (q^{2(c-1)} +\dots+ q^{2(c-d)})t \label{eq:probdens2}\\ 
  &=:& P_x(x)+P_{xt}(x,t).\nonumber
\end{eqnarray}
Note that the time-independent part
\begin{eqnarray*}
  P_x(x)&=&\frac2\pi\left(1-q^{2(s-2)}\right) 
  \sum_{m=0}^\infty q^{2m(s-2)} \frac{1-\cos q^m 2x}{2}\\
  &=& \frac1\pi-\frac{\left(1-q^{2(s-2)}\right)}{\pi} \sum_{m=0}^\infty
  q^{m(2s-4)}\cos q^m 2x, 
\end{eqnarray*}
is a Weierstrass-like function with the dimension
$s'=\max\{2s-2,1\}\in[1,2)$ (i.e.\ for $s\in[1,3/2],\,s'=1$).
From the equation~(\ref{eq:probdens2}) one immediately gets the
spectrum of $P(x,t)$: all the frequencies governing the time evolution
are 
\[
\omega_{c,d} = (q^2-1)(q^{2(c-1)} +\dots+ q^{2(c-d)}),
\]
where $c=1,2,\dots$, $d=1,2,\dots,c$. Thus all the frequencies divide
by $q^2-1$ which is also the smallest frequency, so the fundamental
period of $P(x,t)$ is $2\pi/(q^2-1)$. 

\subsection{\ Average velocity}

Let us study the behavior of $\langle x \rangle$.
\begin{eqnarray*}
  \langle x \rangle & = & \int_0^\pi \!\! dx \,x\,|\Psi|^2\\ 
  & = & \frac\pi2 - \frac{16}{\pi}\left(1-q^{2(s-2)}\right)
  \sum_{k=1}^\infty \frac{q^{k(s-1)}}{(q^{2k}-1)^2} \cos(q^{2k} - 
  1)t. 
\end{eqnarray*}
The above expression is valid only for even $q$. For odd $q$ we have
just the first term, which is $\pi/2$. 

The average $x(t)$ is of class ${\cal C}^1$, because its derivative is
given by an absolutely convergent series:
\begin{eqnarray}
  \left|\frac{d \langle x \rangle}{dt}\right| & = &
  \left|\frac{16}{\pi}\left(1-q^{2(s-2)}\right) \sum_{k=1}^\infty
    \frac{q^{k(s-1)}}{q^{2k}-1} \sin(q^{2k} - 1)t\right| \nonumber \\ 
  & \leq & 
  c \sum_{k=1}^\infty 2 
  \frac{q^{k(s-1)}}{q^{2k}}\nonumber \\ 
  & = & 
  2c \frac{q^{s-3}}{1-q^{s-3}}. \label{avevel}
\end{eqnarray}
We show in section~(\ref{sec:proof}) that~(\ref{avevel}) is fractal,
while for odd $q$ the average velocity $|d \langle x \rangle/dt|$, of
course, is not. This seemingly strange behavior
is caused by the fact that 
\[
\int_0^{\pi} dx \sin nx \sin mx
\]
is non-zero only
for $m,n$ of different parity. 
However, if one
slightly disturbs our function, for instance by changing an arbitrary
number of terms to the next higher or lower eigenstates, the dimensions
$D_x$ and $D_t$ will not be altered, but the average velocity will
become fractal. 
In other words, with probability one, independently of the parity of
$q$, the average velocity of the wave function
\begin{equation}
  \Phi_0(x,t) = M_0 \sum_{n=1}^\infty q^{n(s-2)} \sin((q^n\pm1) x)
  e^{-i(q^n\pm1)^2 t}. 
\end{equation}
is fractal characterized by the same dimensions $D_x$ and $D_t$ as the
function currently studied. 

An explicit example of a similar function for odd $q$ it is  
\begin{equation}
  \label{wellinf2}
  \Phi_1(x,t) = M_1 \left[ 2^{s-2} \sin(2 x) e^{-i2^2t}+
  \sum_{n=1}^\infty q^{n(s-2)} \sin(q^n x) e^{-iq^{2n} 
  t}\right] .
\end{equation}
One can see the only difference between this example and the original
one~(\ref{wellinf}) is in the {\em first\/} term. This difference
accounts for the smoothness or roughness of the average velocity. It
is very interesting because normally one expects that it is the {\em
  asymptotic\/} behavior that determines the fractal dimension. Here
we have exactly opposite case: a change in the first term (varying most
slowly) of a series changes the dimension of a complicated function
$\langle v \rangle$. 

Average velocity of the wave packet~(\ref{wellinf2}) is smooth for
even $q$ and fractal for odd $q$. A function which gives fractal
average velocity for both even and odd $q$ is 
\begin{eqnarray*}
  \Phi_2(x,t) & = & M_2 \left[ 2^{s-2} \sin(2 x) e^{-i2^2t}+
    \sum_{n=0}^\infty q^{n(s-2)} \sin(q^n x) e^{-iq^{2n} 
      t}\right] \\
  & = & M_2 \left[ 2^{s-2} \sin(2 x) e^{-i2^2t}+ \frac1N
    \Psi(x,t)\right],
\end{eqnarray*}
where $M_2$ is the normalization constant. On the other hand,
\[
\Phi_3(x,t) = M_3 \left[ 
  \sum_{n=1}^\infty q^{n(s-2)} \sin(q^n x) e^{-iq^{2n} 
    t}\right] 
\]
gives smooth average velocity for both even and odd $q$.

\bibliography{fractals}
\bibliographystyle{unsrt}

\end{document}